\documentstyle[amsmath,amssymb,aps]{revtex}
\def\nc{\newcommand}
\nc{\nn}{\nonumber}
\nc{\cK}{\cal K}
\nc{\cB}{\cal B}
\nc{\cD}{\cal D}
\nc{\cL}{\cal L}
\nc{\dL}{{\cal L}^{\dagger}}
\nc{\sta}{\sin\theta}
\nc{\cta}{\cos\theta}
\nc{\sda}{\sin^2\theta}
\nc{\cda}{\cos^2\theta}
\nc{\coa}{\cot\theta}
\nc{\sqd}{\sqrt{2}}

\def\p{\partial}
\nc{\pr}{\p_r}
\nc{\pv}{\p_v}
\nc{\pta}{\p_{\theta}}
\nc{\pvi}{\p_{\varphi}}

\nc{\spr}{\p_{r_*}}
\nc{\spv}{\p_{v_*}}
\nc{\spta}{\p_{\theta_*}}
\nc{\spvi}{\p_{\varphi_*}}
\nc{\spdr}{\p^2_{r_*}}
\nc{\spdvr}{\p^2_{r_*v_*}}
\nc{\spdra}{\p^2_{r_*\theta_*}}

\begin{document}

\baselineskip 16pt
\twocolumn[\hsize\textwidth\columnwidth\hsize\csname
           @twocolumnfalse\endcsname

\title{\bf Quantum Thermal Effect of Dirac Particles in a
Non-uniformly Rectilinearly Accelerating Kinnersley Black Hole}
\author{WU Shuang-Qing$^*$ and CAI Xu$^{\dagger}$}
\address{Institute of Particle Physics, Hua-Zhong
Normal University, Wuhan 430079, China \\
\rm $^*$E-mail: sqwu@iopp.ccnu.edu.cn
~~$^{\dagger}$E-mail: xcai@ccnu.edu.cn}
\date{July 7, 2001}
\maketitle

\begin{quote}
\widetext
{\em The Hawking radiation of Dirac particles in an arbitrarily
rectilinearly accelerating Kinnersley black hole is studied by
using a method of the generalized tortoise coordinate transformation.
Both the location and the temperature of the event horizon depend
on the time and polar angle. The Hawking thermal radiation
spectrum of Dirac particles is also derived.

}
\end{quote}
]

\narrowtext
Last decades has witnessed much progress in investigating
the Hawking evaporation$^{\cite{Hawk}}$ of scalar fields
or Dirac particles in the stationary axisymmetry black
holes.$^{\cite{Xuetc,WC}}$ To study the Hawking radiation
of a non-stationary black hole, Zhao and Dai$^{\cite{ZD}}$
proposed an effective method called by the generalized
tortoise coordinate transformation (GTCT). This method
has been applied to discuss quantum thermal effect of
scalar particles in some non-uniformly accelerating
black holes$^{\cite{Zhetc}}$ and that of Dirac particles
in the non-static black hole.$^{\cite{LZ}}$ However, it is
very difficult to study the quantum thermal effect of
Dirac particles in a non-spherically symmetric and
non-stationary black hole. The difficulty is due to
the non-separability of the Chandrasekhar-Dirac
equation$^{\cite{Ch}}$ in the most general space-times.

Recently we succeed in dealing with the Hawking radiation of
Dirac particles in a variable-mass Kerr black hole.$^{\cite{WuC}}$
In this letter, we extend this treatment to the Hawking effect
of Dirac particles in a non-uniformly rectilinearly accelerating
Kinnersley black hole.$^{\cite{Kin}}$ With the aid of the GTCT
method, we study the asymptotic behaviors of first-order and
second-order forms of Dirac equation near the event horizon. We
present the equation that determines the location of the event
horizon of the Kinnersley black hole, and show that both the shape
and the Hawking temperature of the event horizon of Kinnersley black
hole depend on not only the time, but also on the angle. The location
and the temperature coincide with those obtained by investigating
the Hawking effect of Klein-Gordon particles in the accelerating
Kinnersley black hole.$^{\cite{Zhetc}}$

The metric of a non-uniformly rectilinearly accelerating
Kinnersley black hole$^{\cite{Kin}}$ is given in the
advanced Eddington-Finkelstein coordinate system by
\begin{eqnarray}
&&ds^2 = 2dv(Gdv -dr -r^2f d\theta) \nn\\
&&~~~~~ -r^2(d\theta^2 +\sin^2\theta d\varphi^2) \, ,
\end{eqnarray}
where $2G = 1 -2M/r -2a r\cos\theta -r^2f^2$,
$f = -a\sta$. The parameter $a = a(v)$ is the magnitude
of acceleration, the mass $M(v)$ of the hole is a function
of the advanced time $v$.

We choose such a complex null-tetrad in the Kinnersley
black hole that its directional derivatives can be written
as $D = -\pr$, $\Delta = \pv +G\pr$, $\delta = \frac{1}{\sqd r}
\big(-r^2f \pr +\pta +\frac{i}{\sta}\pvi\big)$, and
$\overline{\delta}$ the complex conjugate of $\delta$.
It is not difficult to determine the non-vanishing Newman-Penrose
complex spin coefficients$^{\cite{NP}}$ in the above null-tetrad
as follows
\begin{eqnarray}
&&\rho = 1/r \, , ~~\tilde{\mu} = G/r \, ,
~~\gamma = -G_{,r}/2 = -dG/2dr \, , \nn\\
&&\tau = -\tilde{\pi} = f/\sqd \, ,
~~\beta = \coa/(2\sqd r) \, ,
~~\alpha =\tau -\beta \, , \nn\\
&&\nu = \big[(2rG -r^2G_{,r})f +r^2f_{,v}
+G_{,\theta} \big]/(\sqd r) \, .
\end{eqnarray}

If the back reaction is neglected, the dynamic behavior of spin-$1/2$
test particles in a fixed background spacetime (1) which is of Petrov
type-D is described by the spinor form of the four coupled
Chandrasekhar-Dirac equations$^{\cite{Ch}}$ in the Newman-Penrose
formalism,$^{\cite{NP}}$ namely
\begin{eqnarray}
&&(D +\epsilon -\rho)F_1 +(\overline{\delta} +\tilde{\pi}
-\alpha)F_2 = i\mu G_1/\sqd \, , \nn\\
&&(\Delta +\tilde{\mu} -\gamma)F_2 +(\delta +\beta -\tau)F_1
= i\mu G_2/\sqd \, , \nn\\
&&(D +\epsilon^* -\rho^*)G_2 -(\delta +\tilde{\pi}^*
-\alpha^*)G_1 = i\mu F_2/\sqd \, , \nn\\
&&(\Delta +\tilde{\mu}^* -\gamma^*)G_1 -(\overline{\delta}
+\beta^* -\tau^*)G_2 = i\mu F_1/\sqd \, ,
\end{eqnarray}
where $\mu$ is the mass of Dirac particles. Inserting for the
appropriate spin coefficients into the above equations
and making substitutions $P_1 = \sqd rF_1$, $P_2 = F_2$,
$Q_1 = G_1$, $Q_2 = \sqd rG_2$, we obtain
\begin{eqnarray}
&&-{\cD}_0 P_1 +\big({\cL} -r^2f{\cD}_2\big) P_2 = i\mu rQ_1 \, , \nn\\
&&2r^2{\cB} P_2 +\big({\dL} -r^2f{\cD}_0\big) P_1 = i\mu rQ_2 \, , \nn\\
&&-{\cD}_0 Q_2 -\big({\dL} -r^2f{\cD}_2\big) Q_1 = i\mu rP_2 \, , \nn\\
&&2r^2{\cB} Q_1 -\big({\cL} -r^2f{\cD}_0\big) Q_2 = i\mu r P_1 \, .
\label{reDP}
\end{eqnarray}
in which we have defined operators
${\cB} = \pv +G{\cD}_1 +G_{,r}/2$, ${\cD}_n = \pr +n/r$,
${\cL} = \pta +\frac{1}{2}\coa -\frac{i}{\sta}\pvi$, and
${\dL} = \pta +\frac{1}{2}\coa +\frac{i}{\sta}\pvi$.

Although Eq. (\ref{reDP}) can not be decoupled, to investigate
the Hawking radiation of spin-$1/2$ particles, one should 
concern about the behavior of Dirac spinor components near
the event horizon only. Because the Chandrasekhar-Dirac equation
(\ref{reDP}) can be satisfied by identifying $Q_1$, $Q_2$ with
$\bar{P}_2$, $-\bar{P}_1$, respectively, so we only need deal
with a pair of components $P_1$, $P_2$. As the space-time is
symmetric about $\varphi$-axis, we can introduce the following
GTCT$^{\cite{ZD}}$
\begin{eqnarray}
&&r_* = r +\frac{1}{2\kappa}\ln(r -r_H) \, , \nn\\
&&v_* = v -v_0 \, , ~~ \theta_* = \theta -\theta_0 \, ,
\label{trans}
\end{eqnarray}
where $r_H = r_H(v,\theta)$ is the location of the event horizon,
$\kappa$ is an adjustable parameter and is unchanged under tortoise
transformation. Both parameters $v_0$ and $\theta_0$ are arbitrary
constants.

Under the transformation (\ref{trans}), Eq. (\ref{reDP}) with
respect to ($P_1,P_2$) can be reduced to the following limiting
form near the event horizon
\begin{eqnarray}
&&2r_H^2\big[G(r_H) -r_{H,v}\big]\spr P_2
-\big(r_{H,\theta} +r_H^2f_0\big)\spr P_1  = 0 \, , \nn\\
&&\spr P_1 +\big(r_{H,\theta} +r_H^2f_0 \big)\spr P_2 = 0 \, ,
\label{trDPP}
\end{eqnarray}
after being taken the $r \rightarrow r_H(v_0,\theta_0)$,
$v \rightarrow v_0$ and $\theta \rightarrow \theta_0$ limits.
We denote $f_0 = -a(v_0)\sin\theta_0$.

If the derivatives $\spr P_1$ and $\spr P_2$ in Eq. (\ref{trDPP})
are not equal to zero, the existence condition of nontrial solutions
for $P_1$ and $P_2$ is that the determinant of Eq. (\ref{trDPP})
vanishes, which gives the following equation to determine the
location of horizon
\begin{equation}
2G(r_H) -2r_{H,v} +r_H^2f_0^2 +2f_0 r_{H,\theta}
+r_{H,\theta}^2r_H^{-2} = 0 \, . \label{loca}
\end{equation}
The location of the event horizon and the shape of the black hole
change with time.

Now let us consider the asymptotic behaviors of the second-order
form of Dirac equation near the event horizon. Given the GTCT in
Eq. (\ref{trans}), the limiting form of the second-order equations
for the two-component spinor ($P_1, P_2$), when $r$ approaches
$r_H(v_0, \theta_0)$, $v$ goes to $v_0$ and $\theta$ goes to
$\theta_0$, reads
\begin{eqnarray}
&&{\cK}P_1 +\big[-A +r_H^2G_{,r}(r_H) +r_H^3f_0^2 -r_H^2f_0\coa_0 \nn\\
&&~~-r_H^2f_{0,\theta} -(r_Hf_0 +\coa_0)r_{H,\theta}
-r_{H,\theta\theta}\big] \spr P_1 \nn\\
&&~~+2r_H^2\big\{r_H^2f_{0,v} +G_{,\theta}(r_H)
-G(r_H) r_{H,\theta}r_H^{-1} \nn\\
&&~~-r_Hf_0[r_HG_{,r}(r_H)
+r_{H,v} -2G(r_H)]\big\} \spr P_2 = 0 \, , \label{wone+}
\end{eqnarray}
and
\begin{eqnarray}
&&{\cK}P_2 +\big\{-A +3r_H^2G_{,r}(r_H) +2r_H[2G(r_H) -r_{H,v}] \nn \\
&&~~+5r_H^3f_0^2 -r_H^2f_0\coa_0 +(3f_0 r_H -\coa_0)r_{H,\theta} \nn \\
&&~~-r_H^2f_{0,\theta} -r_{H,\theta\theta}\big\} \spr P_2
+r_{H,\theta}r_H^{-1} \spr P_1 = 0 \, . \label{wone-}
\end{eqnarray}
where the operator ${\cK}$ represents a term involving
the second derivatives
\begin{eqnarray*}
&&{\cK} = \big\{\frac{A}{2\kappa} +2r_H^2[2G(r_H)
-r_{H,v}] +2r_H^4f_0^2 \\
&&~~+2f_0 r_{H,\theta}r_H^2\big\} \spdr
+2r_H^2 \spdvr -2(f_0 r_H^2 +r_{H,\theta}) \spdra
\end{eqnarray*}

With the aid of the event horizon equation (\ref{loca}),
we know that the coefficient $A$ is an infinite limit of
$0/0$-type. By means of the L' H\^{o}spital rule, we get
the following result
\begin{eqnarray}
A &=& \lim_{r \rightarrow r_H} \frac{2r^2(G -r_{H,v})
+r^4f^2 +2f r^2r_{H,\theta} +r_{H,\theta}^2}{r -r_H} \nn\\
&=& 2r_H^2G_{,r}(r_H) +2r_H^3f_0^2 -2r_{H,\theta}^2/r_H \, .
\end{eqnarray}

Now we select the adjustable temperature parameter
$\kappa$ in the operator ${\cK}$ such that
\begin{eqnarray}
r_H^2 &&\equiv \frac{A}{2\kappa} +2r_H^2[2G(r_H) -r_{H,v}]
+2r_H^4f_0^2 \nn\\
&& +2f_0 r_H^2r_{H,\theta} = \frac{r_H^3G_{,r}(r_H)
+r_H^4f_0^2 -r_{H,\theta}^2}{\kappa r_H} \nn\\
&&+2G(r_H)r_H^2 +r_H^4f_0^2 -r_{H,\theta}^2 \, ,
\end{eqnarray}
which means the temperature of the horizon is
\begin{equation}
\kappa =\frac{r_H^2G_{,r}(r_H) +r_H^3f_0^2
-r_{H,\theta}^2/r_H}{r_H^2[1 -2G(r_H)]
-r_H^4f_0^2 +r_{H,\theta}^2} \, . \label{temp}
\end{equation}

With such a parameter adjustment and using Eq. (\ref{trDPP}),
we can recast Eqs. (\ref{wone+},\ref{wone-}) into an united
standard wave equation near the event horizon
\begin{equation}
\big(\spdr +2\spdvr -2C_1 \spdra +2C_2 \spr\big) \Psi = 0 \, ,
\label{wave}
\end{equation}
where $C_1 = f_0 +r_{H,\theta}/r_H^2$ and $C_2$ will be
regarded as finite real constants,
\begin{eqnarray*}
2C_2 &=&2G(r_H)r_H^{-1} -G_{,r}(r_H) -2f_0\coa_0
+2r_{H,\theta}^2r_H^{-3} \\
&-&(\coa_0r_{H,\theta} +r_{H,\theta\theta})r_H^{-2}
+(f_0 +r_{H,\theta}r_H^{-2})\big\{r_H^2f_{0,v} \\
&+&[2r_HG(r_H)-r_H^2G_{,r}(r_H)]f_0 +G(r_H) r_{H,\theta} \\
&-&r_HG_{,\theta}(r_H)\big\}/[G(r_H) -r_{H,v}] \, ,
~~{\rm for}~ \Psi = P_1\, ,\\
2C_2 &=&2G(r_H)r_H^{-1} +G_{,r}(r_H) +2r_Hf_0^2 -2f_0\coa_0  \\
&-&(\coa_0r_{H,\theta} +r_{H,\theta\theta})r_H^{-2}  \, ,
~~{\rm for}~ \Psi = P_2 \, .
\end{eqnarray*}

Separating variables as $\Psi = R(r_*)\exp[\lambda \theta_*
+i(m\varphi -\omega v_*)]$, one obtains
\begin{equation}
R^{\prime\prime} = 2(i\omega -C_0) R^{\prime} \, ,
~~R = R_1 e^{2(i\omega -C_0)r_*} +R_0 \, ,
\end{equation}
where $C_0 = C_2 -\lambda C_1$, in which a real constant
$\lambda$ is introduced in the separation of variables.

The ingoing wave and the outgoing wave to Eq. (\ref{wave}) are
\begin{eqnarray}
&&\Psi_{\rm in} \sim \exp[-i\omega v_* +im\varphi
+\lambda \theta_*] \, ,\\
&&\Psi_{\rm out} \sim \Psi_{\rm in}e^{2(i\omega -C_0)r_*} \, ,
~~ (r > r_H) \, .
\end{eqnarray}

The outgoing wave $\Psi_{\rm out}(r > r_H)$ is not analytic
at the event horizon $r = r_H$, but can be analytically
extended from the outside of the hole into the inside of
the hole through the lower half complex $r$-plane
by $(r -r_H) \rightarrow (r_H -r)e^{-i\pi}$ to
\begin{equation}
\widetilde{\Psi_{\rm out}} = \Psi_{\rm out}
e^{i\pi C_0/\kappa}e^{\pi\omega /\kappa} \, ,~~(r < r_H) \, .
\end{equation}

Following the method of Damour-Ruffini-Sannan's,$^{\cite{DRS}}$
the relative scattering probability of the outgoing wave at the
horizon and the thermal radiation spectrum of Dirac particles from
the event horizon of the hole are
\begin{eqnarray}
&&\left|\frac{{\Psi}_{\rm out}}{\widetilde{\Psi_{\rm out}}}\right|^2
= e^{-2\pi\omega/\kappa} \, , \\
&&\langle {\cal N}(\omega) \rangle
\sim \frac{1}{e^{\omega/T_H} +1} \, , \label{sptr}
\end{eqnarray}
with the Hawking temperature being
\begin{equation}
T_H = \frac{\kappa}{2\pi}
= \frac{1}{4\pi r_H} \cdot \frac{M r_H -r_H^3a\cta_0
-r_{H,\theta}^2}{M r_H +r_H^3a\cta_0 +r_{H,\theta}^2/2} \, .
\end{equation}

In conclusion, we have studied the Hawking radiation of
Dirac particles in an arbitrarily accelerating Kinnersley
black hole whose mass changes with time. Equations (\ref{loca})
and (\ref{temp}) give the location and the temperature of
event horizon of the accelerating Kinnersley black hole,
which depend not only on the advanced time $v$ but also
on the polar angle $\theta$. Eq. (\ref{sptr}) shows the
thermal radiation spectrum of Dirac particles in an
arbitrarily rectilinearly accelerating Kinnersley black
hole. They are in accord with that derived from discussing
on the thermal radiation of Klein-Gordon particles in
the same space-time.$^{\cite{Zhetc}}$

S.Q. Wu is very grateful to Dr. Jeff Zhao at Motorola Company
for his long-term help. This work was supported partially by
the NSFC under Grant No.19875019.

\end{document}